\documentclass[aoas,nameyear,MSNbibl,dvips]{arximspdf}
\usepackage{breakurl}

\doi{10.1214/10-AOAS398L}
\referstodoi{10.1214/10-AOAS398}
\volume{5}
\issue{1}
\pubyear{2011}
\firstpage{56}
\lastpage{60}

\makeatletter

\def\@bmisc[#1]{%
  \get@battribute{unstr}%
  \common@pub@types%
  \let\bauthor\bbl@bauthor%
  \let\bhowpublished\@firstofone%
  \def\borganization##1{{\bauthor@style ##1}}%
}

\makeatother

\begin{document}
\begin{frontmatter}

\title{Discussion of: A statistical analysis of multiple temperature
proxies: Are
reconstructions of surface temperatures over the last\\ 1000 years
reliable?}
\runtitle{Discussion}
\pdftitle{Discussion on A statistical analysis of multiple temperature
proxies: Are reconstructions of surface temperatures over the last 1000
years reliable?
by B. B. McShane and A. J. Wyner}

\begin{aug}
\author[A]{\fnms{Stephen} \snm{McIntyre}\corref{}\ead[label=e1]{stephen.mcintyre@utoronto.ca}}
and
\author[B]{\fnms{Ross} \snm{McKitrick}}

\runauthor{S. M\textup{c}Intyre and R. M\textup{c}Kitrick}

\affiliation{Climate Audit and University of Guelph}

\address[A]{25 Playter Bl\\
Toronto, Ontario\\
M4K2W1, Canada\\
\printead{e1}} 
\end{aug}

\received{\smonth{9} \syear{2010}}
\revised{\smonth{9} \syear{2010}}



\end{frontmatter}

As McShane and Wyner (\citeyear{MW2011}) observe, there are formidable statistical
problems in ``reconstructing'' past temperatures from networks of
so-called ``proxy'' data with weak ``signal'' and complicated
autocorrelated structures. Wegman, Scott and Said (\citeyear{WSS2006}) regretted the combination
of the lack of involvement of the statistical community with the
statistical inexperience of paleoclimatologists struggling with these
complicated problems. Oxburgh et al. (\citeyear{Oetal2010}), one of the Climategate
inquiries, made similar observations. Thus, the interest of McShane and
Wyner in these problems is very much to be welcomed and hopefully
presages wider participation by the statistical community in the very
interesting problems presented by paleoclimate reconstructions.

\section*{Pseudoproxies and benchmarking}

In Sections 2 and 3, MW make a variety of interesting and useful
comparisons between holdout RMSE from a proxy reconstruction using the
lasso method, as against a variety of null models. MW do not translate
their results into ``local'' paleoclimate terminology, which may cause
many paleoclimate scientists to miss or misinterpret some provocative MW
results.

In paleoclimate terms, their results are best interpreted as an extended
commentary on what paleoclimate scientists call the ``RE'' statistic,
the most prevalent statistical test in present-day paleoclimate. In MW
terminology, the RE statistic is calculated from the ratio of the
holdout RMSE from the proxy reconstruction compared to the holdout RMSE
from the in-sample mean:
\begin{eqnarray}
\quad \mathrm{RE}\_\{\mbox{proxy}\} &=& 1 -
\mbox{Holdout}\_\mbox{RMSE}\_\{\mbox{proxy}\}\nonumber\\[-8pt]\\[-8pt]
&&\hphantom{1 -{}}{}/\mbox{Holdout}\_\mbox{RMSE}\_\{\mbox{intercept}\}.\nonumber
\end{eqnarray}
The significance of the RE statistic for a proxy reconstruction is
assessed by comparison with the 95th (99th) percentile of
RE statistics generated from pseudoproxy simulations:
\begin{eqnarray}
\mathrm{RE}\_\{\mbox{pseudoproxy}\} &=& 1 -
\mbox{Holdout}\_\mbox{RMSE}\_\{\mbox{pseudoproxy}\}\nonumber\\[-8pt]\\[-8pt]
&&\hphantom{1 -{}}{}/\mbox{Holdout}\_\mbox{RMSE}\_\{\mbox{intercept}\}.\nonumber
\end{eqnarray}

The boxplots shown in MW Figure 9 and 10 can be readily seen to be
addressing the same issue from a different perspective, with MW
extending\vadjust{\goodbreak} and sharpening previous discussion. Their low-order AR1
pseudoproxy networks implement typical climate science methodology
[e.g., Mann, Bradley and Hughes (\citeyear{MBH1998}); Wahl and Ammann (\citeyear{WA2007})],
in which proxies are
assumed to be well modeled by low order AR1 processes, an assumption
seemingly contradicted by the observed distribution of AR1 coefficients,
as shown, for example, in MW Figure 4 for the network of Mann et al.
(\citeyear{Metal2008}). Their ``empirical AR1'' network implements (and simplifies) the
opposed approach of McIntyre and McKitrick (\citeyear{MM2005a}, \citeyear{MM2005c}), in which
pseudoproxies are constructed to have autocorrelation properties
observed in the actual proxies. Their holdout strategy generalizes
standard climate science practice, by considering all possible holdouts
of 30 years, instead of restricting their holdout to the first and last
half of the reconstruction.

Several of their results are new and surprising. For example, they
observe (see Figure 10) that the performance of proxy reconstructions
with holdout periods at either endpoint (the usual paleoclimate
practice) is noticeably superior to results from interior holdouts. They
also observe (see Figure~9) that white noise pseudoproxies outperform
low order AR1 pseudoproxies (the usual paleoclimate practice). Both
results warrant further investigation.

That empirical AR1 pseudoproxies perform at least as well as proxy
reconstructions is implicit in the RE benchmarks of McIntyre and
McKitrick (\citeyear{MM2005a}, \citeyear{MM2005c}) for the MBH98 network. MW demonstrate that empirical
AR1pseudoproxies outperform actual proxies for the Mann et al. (\citeyear{Metal2008})
network; the degree of outperformance is a surprise.

We entirely agree with MW's conclusion that ``Lasso generated
reconstructions using the proxies\ldots\ do not achieve statistical
significance against sophisticated null models.''

\section*{Lasso}

Some climate scientists will undoubtedly criticize MW for their use of
the Lasso method as a template for comparing proxy and pseudoproxy
networks (as opposed to carrying out the same analysis using the
presently popular paleoclimate methods of CPS and RegEM).

However, for their stated purpose of understanding the statistical
properties of the proxy network, it seems to us that there is much to
recommend the use of a statistical method whose properties are
relatively well understood and which has a relatively efficient
algorithm (as opposed to RegEM), as this enables a focus on the
properties of the proxy network relative to pseudoproxies rather than
intricacies of a nonstandard methodology. We would be very surprised if
their key comparisons of proxy and pseudoproxy results were sensitive to
such methodological variations though the point would be worth checking.

Some climate scientists will probably contest the use of ``empirical''
AR1 coefficients for the construction of pseudoproxy networks---a point
anticipated by MW, as this issue was raised previously by Ammann and
Wahl (\citeyear{AW2007}) against our use of networks constructed using empirical
autocorrelation coefficients. MW disagree sharply with the objection
previously raised by Ammann and Wahl (\citeyear{AW2007}): that using autocorrelation
coefficients estimated from actual proxies results in ``train[ing] the
stochastic engine with significant (if not dominant) low frequency
climate signal rather than purely nonclimatic noise and its
persistence.'' In our opinion, if the proxy networks contained a
``dominant'' or even ``significant'' ``low frequency climate signal''
(as Ammann and Wahl assert but do not demonstrate), then the graphs of
the proxy series would have a consistent low frequency appearance (as
opposed to the visually inconsistent appearance shown in MW Figure 6 and
elsewhere). The very inconsistency of the series within proxy networks
such as Mann et al. (\citeyear{Metal2008}) argues forcefully against the interpretation of
high empirical autocorrelation coefficients as being imported from a
climate ``signal,'' as opposed to being an inherent feature of the
proxies themselves. MW makes the following additional and reasonable
observation in dismissing Ammann and Wahl's objection to empirical AR1
coefficients:

\begin{quote}
it is hard to argue that a procedure is truly skillful if it cannot
consistently outperform noise --- no matter how artfully structured.
\end{quote}

\section*{Reconstructions}

In their Sections 4.2 and 5, MW make temperature reconstructions using
principal components regression on the network of Mann et al. (\citeyear{Metal2008})
proxies extending back to AD1000, this time retaining the first one,
five, ten and 20 principal components of the proxies (after removing
three Tiljander proxies from the same site to avoid singularity). They
observe that remarkably different appearing reconstructions can have
very similar cross-validation statistics, a~phenomenon that very much
complicates the uncertainty. A similar phenomenon can be seen in the
context of a very different network, also involving the use of principal
components, in Briffa et al. (\citeyear{Betal2001}), Figure 4, which presents eight very
different backcasts with virtually indistinguishable cross-validation
statistics. The problem was also raised in McIntyre (\citeyear{M2006}), observing
that a reconstruction could not be ``99.8\% significant'' if there was
an alternative reconstruction with virtually identical cross-validation
properties, but very different backcast medieval values. MW correctly
place the issue squarely back on the table.

Within the family of backcasts, MW observe that the reconstruction with
one PC ``corresponds quite well to backcasts such as those in Mann, Bradley and Hughes
(\citeyear{MBH1999}).'' Not only does this reconstruction ``corresponds quite
well'' to the MBH99 reconstruction; for practical purposes, it pretty
much is MBH99 reconstruction, so the resemblance is unsurprising. There
are no fewer than 19 Graybill strip bark chronologies (mostly
bristlecone) in the 93-series AD1000 network of Mann et al. (\citeyear{Metal2008}).
Because the Graybill bristlecones have a strong common pattern, the
Graybill bristlecones dominate the weights of the PC1 even without the
use of a skew PC methodology [a point made on an earlier occasion in
McIntyre and McKitrick (\citeyear{MM2005b})]. Thus, it is little surprise that the MW
one-PC resembles the MBH99 reconstruction, since both reconstructions
are weighted heavily toward the same Graybill strip bark bristlecone
chronologies.

The tendency of the early portion of the MW reconstructions to increase
with more PCs reflects a phenomenon involving bristlecone weighting that
has attracted considerable comment and controversy in many venues. As
more PCs are added to the network, the weight of the bristlecones is
diminished, resulting in a less hockey-stick shaped reconstruction---as
also observed in McIntyre and McKitrick (\citeyear{MM2005a}, \citeyear{MM2005b}).

MW observe that the standard methods of estimating uncertainty in
paleoclimate literature do not remotely address the underlying
complications of the multivariate methodology. Their own estimates of
uncertainty are much wider than the uncertainty bands in Mann et al.
(\citeyear{Metal2008}). Despite these very large increases, it is not clear to us that
even these wider bands fully allow for the problem of proxy
inconsistency. In our own comment on Mann et al. (\citeyear{Metal2008}) [McIntyre and
McKitrick (\citeyear{MM2009})], we observed that paleoclimate reconstructions are an
application of multivariate calibration and that the inconsistency test
of Brown and Sundberg (\citeyear{BS1987}) applied to the AD1000 network of Mann et
al. (\citeyear{Metal2008}) yielded infinite confidence intervals prior to AD1800. The
difference between these results and the MW estimates warrants close
examination.

It also needs to be clearly recognized, that, even though MW's results
are rather discouraging for the reconstructions using the Mann et al.
(\citeyear{Metal2008}) network, they are, in a sense, a \textit{best case} as they
\textit{assume} that the quality of the data set is satisfactory
(thereby not taking a position on prominent controversies over the
proxies within this data set). For example, the Korttajarvi sediment
series have been contaminated in their modern portion by bridge and
other construction sediments, a point made in the original publication
[Tiljander et al. (\citeyear{Tetal2003})]. The explosive increase in these series is due
to nonclimatic causes, contrary to the best case assumption stipulated
by MW. The correlation of this nonclimatic increase with temperature is
a classic example of spurious correlation, one which, in this case,
leads ironically to a reversal of the data set from the orientation set
out in the underlying publication.

Similarly, although Mann et al. (\citeyear{Metal2008}) stated that they would adhere to
recommendations of the 2006 NAS panel, they did not comply with
recommendations by the panel that strip bark tree ring chronologies
(i.e., bristlecones, foxtails) be ``avoided'' in temperature
reconstructions, using no fewer than 23 of the most disputed strip bark
chronologies.


\printaddresses

\end{document}